\documentclass[aps,prd,twocolumn,groupedaddress,showpacs,nofootinbib,amssymb]{revtex4}
\usepackage[T1]{fontenc}
\usepackage[latin1]{inputenc}
\usepackage{graphicx}
\usepackage[english]{babel}
\usepackage{amsmath}
\usepackage{amssymb}
\usepackage{amsfonts}

\begin{document}

\def\pp{{\, \mid \hskip -1.5mm =}}
\def\cL{{\cal L}} 
\def\beq{\begin{equation}}
\def\eneq{\end{equation}}
\def\bea{\begin{eqnarray}}
\def\enea{\end{eqnarray}}
\def\tr{{\rm tr}\, }
\def\nn{\nonumber \\}
\def\e{{\rm e}}

\title{\textbf{O'Hanlon actions  by Noether symmetry}}

\author{F. Darabi$^\ast$}

\affiliation{\it $^\ast$Department of Physics, Azarbaijan Shahid Madani University, Tabriz 53741-161, Iran\\
Research Institute for Astronomy and Astrophysics of Maragha (RIAAM), Maragha 55134-441, Iran}

\date{\today}

\begin{abstract}
By using the conformal symmetry between Brans-Dicke action with $\omega=-\frac{3}{2}$ and O'Hanlon action, we seek the O'Hanlon actions in Einstein frame respecting the Noether symmetry. Since the Noether symmetry is preserved under conformal transformations, the existence of Noether symmetry in the Brans-Dicke action asserts the Noether symmetry in O'Hanlon action in Einstein frame. Therefore, the potentials respecting Noether symmetry in Brans-Dicke action give the corresponding potentials respecting Noether symmetry in O'Hanlon action in Einstein frame. 

\end{abstract}
\maketitle

\section{introduction}
\label{1}

Alternative theories of gravity (general relativity) have been formulated and investigated in different contexts. The most former theory in this ground
is known as Brans-Dicke scalar-tensor gravity in which the coupling of the scalar field to the geometry is nonstandard \cite{Brans} so that the gravitational coupling turns out to be no longer constant. Since then, more general couplings were considered and the compatibility of such approaches with the different formulations of equivalence principle have been studied \cite{Dicke-Penrose}-\cite{Dicke-Penrose11}. By conformal transformations it is possible to show that in the absence of ordinary matter, any scalar-tensor theory within the Jordan approach \cite{Jordan},
\cite{Jordan1} is conformally equivalent to an Einstein theory within the Einstein approach, plus a minimally coupled scalar field \cite{Capozziello},
\cite{Capozziello1}. Here, we show that such a conformal transformation when applied to the nonstandard Brans-Dicke action with the special parameter $\omega=-\frac{3}{2}$ leads to the O'Hanlon type theory \cite{Ohanlon} where the kinetic term is removed, i.e. the dynamics is completely endowed by the self interacting potential. On the other hand, it is generally shown that the Noether symmetry is preserved under the conformal transformation \cite{Capozziello},
\cite{Capozziello1}. Therefore, if we find that the Noether symmetry exists in such Brans-Dicke action, then we may conclude that the
Noether symmetry exists in O'Hanlon action, too. Fortunately, such a Noether symmetry has already been obtained in Brans-Dicke action for a specific scalar field potential \cite{Roshan}. Hence, we use this potential in Brans-Dicke action and obtain the corresponding potential which respects the Noether symmetry in the O'Hanlon action. Our motivation for this procedure is the fact that O'Hanlon action has no dynamical term, so it is not easy to apply the Noether symmetry approach directly to this action.  
\\ In Sec. (\ref{2}) we discuss the conformal symmetry between Jordan and Einstein frames. In Sec. (\ref{3}) we introduce the conformal transformation
which transforms the Brans-Dicke action with $\omega=-\frac{3}{2}$ into O'Hanlon action, and in Sec. (\ref{4}) we apply the Noether symmetry method to obtain
the self interacting scalar fields which preserve the Noether symmetry in
both theories. Conclusions are given in Sec. (\ref{5}).  

\section{Conformal symmetry between Jordan and Einstein frames} \label{2}

The general form of the action in four dimensions for a nonstandard coupling between the scalar field and the geometry is given by
\begin{equation}\label{eq01}
{\cal S}=\int d^4x \sqrt{-g}\left(F(\phi)R+\frac{1}{2}g^{\mu \nu}\phi_{;\mu}\phi_{;\nu}-V(\phi)\right),
\end{equation}
where $R$ is the Ricci scalar, $V(\phi)$ and $F(\phi)$ are typical functions
describing the potential for the scalar field $\phi$ and the coupling of $\phi$ with gravity, respectively\footnote{The metric signature is $(-+++)$
and Planck units are used.}. This form of the action or the Lagrangian density
is usually referred to the {\it Jordan frame}, because of the coupling term
term $F(\phi)R$. The variation with respect to the metric $g_{\mu
\nu}$ gives rise to the generalized Einstein equations
\begin{equation}\label{eq02}
F(\phi)G_{\mu \nu}=-\frac{1}{2}T_{\mu \nu}-g_{\mu \nu}\square_{\Gamma}F(\phi)+F(\phi)_{;\mu\nu},
\end{equation}
where $\square_{\Gamma}$ is the d'Alembert operator with respect to the connection
$\Gamma$, $G_{\mu \nu}$ is the standard Einstein tensor
\begin{equation}\label{eq03}
G_{\mu \nu}=R_{\mu \nu}-\frac{1}{2}R g_{\mu \nu},
\end{equation}
and $T_{\mu \nu}$ is the energy-momentum tensor of the scalar field
\begin{equation}\label{eq04}
T_{\mu \nu}=\phi_{;\mu}\phi_{;\nu}-\frac{1}{2}g_{\mu \nu}\phi_{;\alpha}\phi^{;\alpha}+g_{\mu \nu}V(\phi).
\end{equation}
The variation with respect to $\phi$ leads to Klein-Gordon equation
\begin{equation}\label{eq05}
\square_{\Gamma}\phi-RF_{\phi}(\phi)+V_{\phi}(\phi)=0,
\end{equation}
where $F_{\phi}=\frac{dF(\phi)}{d\phi}, V_{\phi}(\phi)=\frac{dV(\phi)}{d\phi} $. We now consider a conformal transformation on the metric $g_{\mu \nu}$
\begin{equation}\label{eq06}
\bar{g}_{\mu \nu}=e^{2\Omega}g_{\mu \nu},
\end{equation}
where $\Omega$ is an arbitrary function of spacetime. The Riemann and Ricci tensors together
with the connection and Ricci scalar transform under this conformal transformation
so that the Lagrangian density in (\ref{eq01}) becomes
\bea    \label{eq07}
\sqrt{-g}\left(FR+\frac{1}{2}g^{\mu \nu}\phi_{;\mu}\phi_{;\nu}-V(\phi)\right)=\sqrt{-\bar{g}}e^{-2\Omega}\left(F\bar{R}+ \right.\nonumber \\
\left.-6F\square_{\bar{\Gamma}}\Omega-6F\Omega_{;\alpha}\Omega^{;\alpha}+\frac{1}{2}\bar{g}^{\mu \nu}\phi_{;\mu}\phi_{;\nu}-e^{-2\Omega}V(\phi)\right),
\enea
where $\bar{R}, \bar{\Gamma}$ and $\square_{\bar{\Gamma}}$ are the corresponding
quantities with respect to the metric $\bar{g}_{\mu \nu}$ and connection
$\bar{\Gamma}$, respectively. If we require the new theory in terms of $\bar{g}_{\mu \nu}$ to appear as a standard Einstein theory the conformal factor has to
be related to $F$ as
\begin{equation}\label{eq08}
e^{2\Omega}=2F.
\end{equation}
Using this relation, the Lagrangian density (\ref{eq07}) becomes
\bea\label{eq09}
&&\sqrt{-g}\left(FR+\frac{1}{2}g^{\mu \nu}\phi_{;\mu}\phi_{;\nu}-V(\phi)\right)\\
\nonumber 
&=&\sqrt{-\bar{g}}\left(\frac{1}{2}\bar{R}+3\square_{\bar{\Gamma}}\Omega+ \frac{3F_{\phi}^2-F}{4F^2}\phi_{;\alpha}\phi^{;\alpha}-\frac{V(\phi)}{4F^2}\right).
\enea
By introducing a new scalar field $\bar{\phi}$ and the potential $\bar{V}$,
respectively defined by
\begin{equation}\label{eq10}
\bar{\phi}_{;\alpha}=\sqrt{\frac{3F_{\phi}^2-F}{4F^2}}\phi_{;\alpha},
\:\:\:\:\: \bar{V}(\bar{\phi}(\phi))=\frac{V(\phi)}{4F^2},
\end{equation}
we obtain\footnote{Note that the divergence-type term $3\square_{\bar{\Gamma}}\Omega$ appearing in the Lagrangian density is not considered \cite{Capozziello},
\cite{Capozziello1}.}
\bea    \label{eq11}
&&\sqrt{-g}\left(FR+\frac{1}{2}g^{\mu \nu}\phi_{;\mu}\phi_{;\nu}-V(\phi)\right)\nonumber \\
&=&\sqrt{-\bar{g}}\left(\frac{1}{2}\bar{R}+\frac{1}{2}\bar{\phi}_{;\alpha}\bar{\phi}^{;\alpha}-\bar{V}(\bar{\phi})\right),
\enea
where the r.h.s. is the usual Einstein-Hilbert Lagrangian density subject to the metric
$\bar{g}_{\mu \nu}$, plus the standard Lagrangian density of the scalar field
$\bar{\phi}$. This form of the Lagrangian density is
usually referred to the {\it Einstein frame}. Therefore, we realize that any nonstandard coupled theory of gravity with scalar field, in the absence of ordinary matter, is conformally equivalent to the standard Einstein gravity coupled with scalar field provided that we use the conformal transformation (\ref{eq08}) together with the definitions (\ref{eq10}).
The converse is also true: for a given $F(\phi)$, such that ${3F_{\phi}^2-F}>0$,
we can transform a standard Einstein theory into a nonstandard coupled theory.
This has an important meaning: if we are able to solve the field equations
within the framework of standard Einstein gravity coupled with an scalar
field subject to a given potential, we should be able to get the solutions
for the class of nonstandard coupled theories, with the coupling $F(\phi)$,
through the conformal transformation and the definitions defined by (\ref{eq08}), and (\ref{eq10}), respectively. This statement is exactly what we mean as the {\it conformal equivalence between Jordan and Einstein frames}.

\section{Conformal symmetry between Brans-Dicke action with $\omega=-\frac{3}{2}$ and O'Hanlon action}\label{3}

The Brans-Dicke action in Jordan frame with $\omega=-\frac{3}{2}$ is defined by the action
\begin{equation}\label{eq27}
{\cal S}=\int_P d^4x \sqrt{-g}\left(\phi
R+\frac{3}{2\phi}\phi_{;\mu} \phi^{;\mu}-V(\phi)\right).
\end{equation}
We are motivated for choosing this action with the specific $\omega$ because it is known that the Brans-Dicke action in Jordan frame with $\omega=-\frac{3}{2}$
is equivalent to the $f({\cal R})$ gravity in Palatini formalism, where ${\cal
R}$ is the Ricci scalar \cite{Sotiriou}. Moreover, the viable $f({\cal R})$ theories of gravity in Palatini formalism were found by searching for
the Noether symmetry within the dynamically equivalent action, namely the Brans-Dicke action in Jordan frame with $\omega=-\frac{3}{2}$, with some viable scalar field potentials  \cite{Roshan}.
Here, by the study of  conformal equivalence between the Brans-Dicke action
in Jordan frame with $\omega=-\frac{3}{2}$ and O'Hanlon action in Einstein
frame we are indeed looking for those O'Hanlon actions in Einstein
frame which are viable regarding the viable $f({\cal R})$ theories of gravity in Palatini formalism, from Noether symmetry point of view.
In this way, for any viable $f({\cal R})$ theory of gravity from Noether symmetry point of view, we may find the
corresponding O'Hanlon actions in Einstein
frame. 
 
By redefining the scalar field $\phi$ to a new field
\begin{equation}\label{eq28}
\sigma=2\sqrt{3\phi},
\end{equation}
the Brans-Dicke action (\ref{eq27}) becomes
\begin{equation}\label{eq29}
{\cal S}=\int_P d^4x \sqrt{-g}\left(F(\sigma)
R+\frac{1}{2}g^{\mu \nu}\sigma_{;\mu}\sigma_{;\nu}-V(\sigma)\right),
\end{equation}
where
\begin{equation}\label{eq30}
F(\sigma)=\frac{1}{12}\sigma^2.
\end{equation}
This action is now exactly the same as (\ref{eq01}) in the Jordan frame in which $\phi$ is replaced by $\sigma$.
Therefore, with a similar procedure for the field $\sigma$ we may write down
\bea\label{eq31}
&&\sqrt{-g}\left(F(\sigma)R+\frac{1}{2}g^{\mu \nu}\sigma_{;\mu}\sigma_{;\nu}-V(\sigma)\right)\nonumber \\
&=&\sqrt{-\bar{g}}\left(\frac{1}{2}\bar{R}+\frac{1}{2}\bar{\sigma}_{;\alpha}\bar{\sigma}^{;\alpha}
-\bar{V}\right),
\enea
where
\begin{equation}\label{eq32}
\bar{\sigma}_{;\alpha}=\sqrt{\frac{3F_{\sigma}^2-F}{4F^2}}\sigma_{;\alpha},
\:\:\:\:\: \bar{V}(\bar{\sigma}(\sigma))=\frac{V(\sigma)}{4F^2},
\end{equation}
and
\beq
F_\sigma=\frac{dF(\sigma)}{d\sigma}.
\eneq
Substituting $F(\sigma)$ (\ref{eq30}) in the definition of $\bar{\sigma}_{;\alpha}$ leads to zero kinetic term for this field and we obtain
\bea\label{eq33}
\sqrt{-\bar{g}}\left(\frac{1}{2}\bar{R}-\bar{V}\right).
\enea
The r.h.s. of eq. (\ref{eq33}) is the Lagrangian density in the Einstein frame \footnote{
It is interesting to note that for the following potential
\begin{equation}\label{eq34}
V(\sigma)=\frac{\bar{\Lambda}}{36}\sigma^4,
\end{equation}
where $\bar{\Lambda}$ is a constant, we obtain $\bar{V}=\bar{\Lambda}$ and the action in Einstein frame is reduced to Einstein-Hilbert action with a cosmological constant $\bar{\Lambda}$. The corresponding potential
in the Jordan frame with Brans-Dicke action (\ref{eq27}) takes the following form
\begin{equation}\label{eq35}
V(\phi)=4\bar{\Lambda}\phi^2,
\end{equation}
which converts the action into a gravity theory non-minimally coupled with
a massive scalar field with an squared mass scale of the order of cosmological
constant.}, namely it becomes the O'Hanlon action where the dynamics is completely endowed by the self interacting potential \cite{Ohanlon}. Therefore, it is shown that the Brans-Dicke action in Jordan frame with the parameter $\omega=-\frac{3}{2}$ and O'Hanlon action in Einstein frame are conformally equivalent.

\section{Noether symmetry in Brans-Dicke action with $\omega=-\frac{3}{2}$}\label{4}

Using the flat Friedmann-Robertson-Walker metric the Lagrangian related to the action (\ref{eq27}) takes the point-like form
\begin{eqnarray}\label{eq36}
{\cal L}=12a^{2}\varphi \dot{\varphi}
\dot{a}+6\varphi^{2}\dot{a}^{2}a+6a^{3}\dot{\varphi}^{2}-V(\varphi)a^{3},
\end{eqnarray}
where the redefinition $\phi\equiv\varphi^{2}$ has been used.
Solutions for the dynamics given by (\ref{eq36}) can be achieved by selecting cyclic variables related to some Noether symmetry \cite{defelice, lambiase}. In principle, this approach allows us to select gravity models compatible with the Noether symmetry.  

Let $\mathcal{L}(q^i,\dot{q}^i)$ be a canonical, non degenerate point-like Lagrangian subject to
\beq
\frac{\partial\mathcal{L}}{\partial t}=0, \hspace{1.5cm}  det H_{ij}\equiv \left\|\frac{\partial^2\mathcal{L}}{\partial \dot{q}^i\partial \dot{q}^j}\right\|\neq0,
\eneq
where $H_{ij}$ is the Hessian matrix and a dot denotes derivative with respect to the cosmic time $\textit{t}$. The Lagrangian $\mathcal{L}$ is generally
of the form
\beq
\mathcal{L}=T(\textbf{q},\dot{\textbf{q}})-V(\textbf{q}),                \eneq
where \textit{T} and \textit{V} are kinetic energy (with positive definite quadratic form) and potential energy, respectively. The energy function associated with $\mathcal{L}$ is defined by
\beq
E_\mathcal{L}\equiv\frac{\partial\mathcal{L}}{\partial \dot{q}^i}-\mathcal{L},
\eneq
which is the total energy $T + V$ as a constant of motion. Since our cosmological problem has a finite number of degrees of freedom, we consider only point transformations.

Any invertible transformation of the generalized positions $Q^i=Q^i(\textbf{q})$ induces a transformation of the generalized velocities 
\beq
\dot{Q}^i(\textbf{q})=\frac{\partial Q^i}{\partial q^j}\dot{q}^j,   \label{4.23}
\eneq
where the matrix $\mathcal{J}=\left\|\partial Q^i/\partial q^j\right\|$ is the Jacobian of the transformation, and it is assumed to be non-zero. On the other hand, an infinitesimal point transformation is represented by a generic vector field on $Q$
\beq
\textbf{X}=\alpha^i(\textbf{q})\frac{\partial}{\partial q^i}.
\eneq
\\ The induced transformation (\ref{4.23}) is then represented by
\beq
\textbf{X}^c=\alpha^i\frac{\partial}{\partial q^i}+\left(\frac{d}{dt}\alpha^i\right)\frac{\partial}{\partial \dot{q}^i}.   \label{4.24}
\eneq
The Lagrangian $\mathcal{L}(\textbf{q}, \dot{\textbf{q}})$ is invariant under the transformation by \textbf{X} provided that
\beq
L_X\mathcal{L}\equiv\alpha^i\frac{\partial \mathcal{L}}{\partial q^i}+\left(\frac{d}{dt}\alpha^i\right)\frac{\partial}{\partial \dot{q}^i}\mathcal{L}=0,
\eneq
where $L_X\mathcal{L}$ is the Lie derivative of ${\mathcal{L}}$.
Let us now consider the Lagrangian $\mathcal{L}$ and its Euler-Lagrange equations
\beq
\frac{d}{dt}\frac{\partial\mathcal{L}}{\partial \dot{q}^j}-\frac{\partial\mathcal{L}}{\partial q^j}=0.                \label{4.25}
\eneq
Contracting (\ref{4.25}) with $\alpha^i$ gives
\beq
\alpha^j\left(\frac{d}{dt}\frac{\partial\mathcal{L}}{\partial \dot{q}^j}\right)=\alpha^j\left(\frac{\partial\mathcal{L}}{\partial q^j}\right).     \label{4.26}      
\eneq
On the other hand, we can write
\beq \label{4.26'}
\frac{d}{dt}\left(\alpha^j\frac{\partial\mathcal{L}}{\partial \dot{q}^j}\right)=\alpha^j\left(\frac{d}{dt}\frac{\partial\mathcal{L}}{\partial \dot{q}^j}\right)+\left(\frac{d\alpha^j}{dt}\right)\frac{\partial\mathcal{L}}{\partial \dot{q}^j},
\eneq
in which the first term in the RHS can be replaced by the RHS of (\ref{4.26}),
hence (\ref{4.26'}) results in
\beq
\frac{d}{dt}\left(\alpha^j\frac{\partial\mathcal{L}}{\partial \dot{q}^j}\right)=L_X\mathcal{L}.
\eneq
The immediate consequence of this result is the \textit{Noether theorem} which states: if $L_X\mathcal{L}=0$, then the function
\beq
\Sigma_0=\alpha^k\frac{\partial\mathcal{L}}{\partial \dot{q}^k},                                                \label{4.27}
\eneq
is a constant of motion. 

In the present model of scalar-tensor cosmology, the Lagrangian is defined by (\ref{eq36}), and the generator of the symmetry corresponding to this Lagrangian is given by
\beq
\textbf{X}=A\frac{\partial}{\partial a}+B\frac{\partial}{\partial \phi}+\dot{A}\frac{\partial}{\partial \dot{a}}+\dot{B}\frac{\partial}{\partial \dot{\phi}}.           \label{4.28}
\eneq
The Noether symmetry exists if the equation $L_X\mathcal{L}=0$ has solution
for the Killing vector $X$. In other words, a symmetry exists if at least one of the functions $A$, or $B$ in the equation (\ref{4.28}) is different from zero.  The existence condition for the symmetry leads to the following system of partial differential equations \cite{Roshan}
\begin{eqnarray}\label{4.39}
2\varphi A+aB+\varphi^{2}\frac{\partial
A}{\partial\varphi}+a\varphi\frac{\partial A}{\partial
a}+a^{2}\frac{\partial B}{\partial a}+a\varphi\frac{\partial
B}{\partial \varphi}=0,
\end{eqnarray}
\begin{eqnarray}\label{4.40}
\varphi A+2aB+2a\varphi\frac{\partial A}{\partial
a}+2a^{2}\frac{\partial B}{\partial a}=0,
\end{eqnarray}
\begin{eqnarray}\label{4.41}
3A+2\varphi\frac{\partial A}{\partial\varphi}+2a\frac{\partial
B}{\partial\varphi}=0,
\end{eqnarray}
\begin{eqnarray}\label{4.42}
3a^{2}V(\varphi)A+B\frac{dV}{d\varphi}a^{3}=0.
\end{eqnarray}
From Eq.(\ref{4.42}) we have
\begin{eqnarray}\label{4.43}
A=\left[-\frac{V'(\varphi)}{3V(\varphi)}\right]Ba,
\end{eqnarray}
where $'$ denotes derivative with respect to $\phi$. Substituting (\ref{4.43}) into (\ref{4.40}), we find that $A=f(\varphi)a^{n}$ and
\beq
\frac{V'}{3V}=\frac{2n}{1+2n} \varphi^{-1}.
\eneq
By substituting these results in (\ref{4.41}) we obtain
\begin{eqnarray}\label{4.44}
f(\varphi)=\beta \varphi^{n-1}
\end{eqnarray}
where $\beta$ is a constant. These results satisfy Eq.(\ref{4.39}) for any arbitrary $n$. From Eqs.(\ref{4.43})
and (\ref{4.44}) one obtains \cite{Roshan}
\begin{eqnarray}\label{4.45}
A=\beta a^{n}\varphi^{n-1}, 
\end{eqnarray}
\begin{eqnarray}\label{4.45'}
B=-\frac{(2n+1)\beta}{2n}a^{n-1}\varphi^{n}
\end{eqnarray}
\begin{eqnarray}\label{4.46}
V(\varphi)=\lambda
\varphi^{\frac{6n}{1+2n}},
\end{eqnarray}
or
\begin{eqnarray}\label{4.46'}
V(\phi)=\lambda
\phi^{\frac{3n}{1+2n}},
\end{eqnarray}
where $\lambda$ is a constant. In conclusion, the Noether symmetry
for the Lagrangian (\ref{eq36}) with the potential (\ref{4.46}) exists and the associated vector field $X$ is determined by (\ref{4.45}) and (\ref{4.45'})
provided that $n\neq 0, -1/2$.
\vspace{10mm}

\section{Noether symmetry in O'Hanlon action }\label{5}

If we apply the Noether symmetry approach for O'Hanlon action we realize
that the corresponding Lagrangian is degenerate and this cause a serious
problem because the symmetric vector field $X$ looses one degree of freedom
related to the velocity of the scalar field. To overcome this problem, we
remind that according to \cite{Capozziello}, \cite{Capozziello1}, Noether symmetries are conformally preserved in a general way. Bearing this in mind, we note that the Noether symmetry in the action (\ref{eq27}) with the potential (\ref{4.46'}) is preserved under a conformal transformation into O'Hanlon action (\ref{eq33}). Therefore,
O'Hanlon action represents the Noether symmetry provided that we have the following class of potentials 
\begin{equation}\label{4.47}
\bar{V}(\sigma)={3\lambda}\sigma^{-\frac{2n+4}{2n+1}},\,\,\,\,\,n\neq 0, -{1}/{2},
\end{equation}
where use has been made of Eqs.(\ref{eq28}), (\ref{eq30}), (\ref{eq32}) and
(\ref{4.46'}).

\section*{Acknowledgment}

This work has been supported financially by Research Institute
for Astronomy and Astrophysics of Maragha (RIAAM) under research project
NO.1/1751.

\end{document}